\documentstyle[graphicx,aps,multicol]{revtex}

\begin{document}
\draft
\title{Intermittency in forced two-dimensional turbulence}
\author{W. Brent Daniel, Maarten A. Rutgers}
\address{Department of Physics, The Ohio State University, 174 W. 18th
Ave. Columbus, OH 43210}
\date{\today}
\maketitle

\begin{abstract}
We find strong evidence for intermittency in forced two dimensional (2D)
turbulence in a flowing soap film experiment.  In the forward enstrophy cascade
the structure function scaling exponents are nearly indistinguishable from 3D
studies. Intermittency corrections are present in the inverse energy cascade as
well, but weaker. Stretched exponential tails of the velocity difference
probability distribution functions and shock like events at large velocity
differences also resemble 3D studies. For decaying turbulence, where only the
forward enstrophy cascade remains, all signs of intermittency disappear.
\end{abstract}


\begin{multicols}{2}

Energy dissipation in three dimensional (3D) turbulence is punctuated by
intermittent bursts \cite{Batchelor49a} and it has been conjectured that
intense vortex filaments are responsible for these bursts\cite{Frisch95a}.
Since vortex filaments cannot exist in two dimensions (2D) it is perhaps not
surprising that recent  experiments by Paret and Tabeling on the inverse energy
and forward enstrophy cascades of 2D turbulence have found no signs of
intermittency\cite{Paret98a,Paret99a}.  Simulations by Smith and Yakhot
\cite{Smith94a} for the inverse energy cascade agree with these experiments,
but simulations by Babiano, Dubrulle, and Frick\cite{Babiano95a} do not.  The
latter simulations go beyond the experiments in that they explore not only the
isolated inverse energy and forward enstrophy cascades but also simultaneous
cascades.  In each case Babiano {\sl et al.} find intermittency only in the
energy cascade, which is partly at odds with the experiments. Further work
could resolve such discrepancies. In this letter we report on 2D turbulence
experiments in flowing soap films which can probe an isolated enstrophy
cascade, as done by Paret and Tabeling\cite{Paret99a}, or simultaneous cascades
as simulated by Babiano {\sl et al.}  The presence of intermittency in 2D would
call for fundamentally new ideas about the source of the phenomenon in 2D and
perhaps even in 3D.

Kraichnan \cite{Kraichnan67a} proposed that there are two scaling regimes in
the energy spectrum of isotropic homogeneous 2D turbulence.   Energy
conservation, accompanied by the assumption that the energy should depend only
on the wavenumber $k$ and the energy dissipation rate per unit mass
$\varepsilon$ leads to, $E(k) = C \varepsilon^{2/3} k^{-5/3}$.   This follows
the same dimensional arguments first applied to 3D turbulence by Kolmogorov
\cite{Kolmogorov41a}. In 2D there is a further constraint. The mean square
vorticity, or enstrophy ($\Omega = 1/2 \; |\nabla \times \mathbf{v}|^{\rm 2}$),
must also be conserved. Through the same considerations as above, namely the
dependence on $k$ and the enstrophy dissipation rate per unit mass $\eta$, we
find $E(k) = C^\prime \eta^{2/3} k^{-3}$. This implies the existence of a
second cascade, the enstrophy cascade. In order for the two cascades to be
present simultaneously, Kraichnan proposed a forward enstrophy cascade and an
inverse energy cascade\cite{Kraichnan67a}.

It is in these two inertial ranges that we look for the presence of
intermittency. A particularly useful tool is the velocity difference, or
increment, between two points separated by a vector $\mathbf{r}$, $\bbox{\delta
v}(\mathbf{x}, \mathbf{r}) = \mathbf{v}(\mathbf{x} + \mathbf{r}) -
\mathbf{v}(\mathbf{x})$. In the present experiment we take both $\mathbf{v}$
and $\mathbf{r}$ along the direction of the mean flow. Using this quantity it
becomes possible to probe the statistics of the flow as a function of length
scale $r$.  If the turbulent velocity field is self similar with respect to
$r$, then the probability distribution functions $P(\delta v(r))$ should scale
with $r$, as was the case in the experiments of Paret and
Tabeling\cite{Paret98a,Paret99a}. Intermittent events break the self similarity
of the flow field and subsequently lead to $P(\delta v(r))$ which do not scale
with $r$.

This deviation from self-similarity is also manifest in the moments of the
velocity differences, collectively known as the structure functions,

\begin{equation}
S_p(r) = \; \langle \delta v^p(r) \rangle \\ =\int_{-\infty}^{\infty} \delta
v^p P(\delta v(r))\,d(\delta v). \label{eq:Sp}
\end{equation}

\noindent We also define a second function, $G_p(r) =  \langle |\delta v(r)|^p
\rangle$, which is often calculated for odd $p$ when a scaling range in
$S_p(r)$ is narrow or absent. Though the theoretical implications of $G_p(r)$,
with $p$ odd, are not yet clear we include it for completeness.  In any case,
results derived from $G_{\rm odd}(r)$ follow exactly the same trend as those
derived from $S_{\rm even}(r) \equiv G_{\rm even}(r)$.

The only exact result involving the structure functions is the 2D equivalent of
Kolmogorov's four-fifths law, relating the third order structure function to
the energy dissipation rate \cite{Paret98a}. It gives, $S_3(r) =
\frac{3}{2}\varepsilon r$. Though no analytic relation has been derived for
structure functions of higher orders, self-similarity implies that the
structure functions should scale like $S_p(r) \propto r^{\zeta_p}$, with
$\zeta_p = p/3$ in the energy cascade. Dimensional analysis also suggests that
$S_3(r) \propto r^3$ and $\zeta_p = p$ in the enstrophy cascade
\cite{Benzi86a}. However, because we will look at the scaling exponents using
the extended self-similarity (ESS) \cite{Benzi93a} hypothesis in the analysis
of the structure functions, the results expected in the absence of
intermittency will be the same in the enstrophy cascade as those in the energy
cascade.

\begin{figure}[tbp] \includegraphics[width=3.2in]{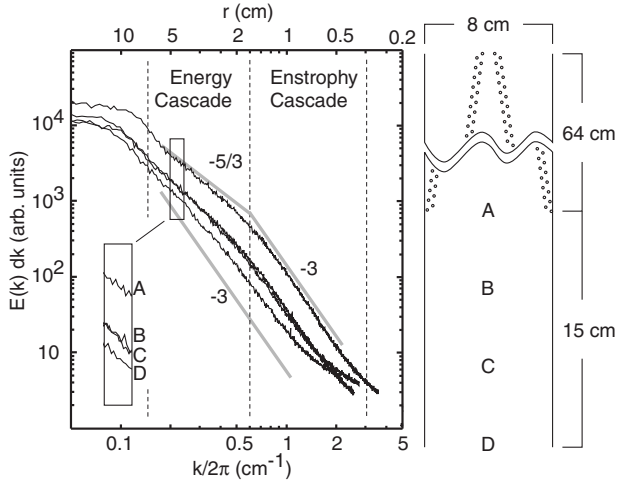}
\vspace{0.1in} \caption{Energy spectra at locations A though D as noted at
right. The energy and enstrophy cascade ranges are with respect to curve A, the
fully forced case. Theoretically predicted energy and enstrophy scaling
exponents are provided for comparison with the data.} \label{fig:fouriers}
\end{figure}

The present experiment makes measurements on a flowing soap film suspended
between two taut vertical nylon wires using an apparatus similar to that of
previous experiments\cite{Rutgers98a}. Two combs, each 64 cm long, are placed
in the film so that they form an inverted wedge 1.2 cm apart at the top and 8
cm apart at the bottom. The cylindrical teeth have an average spacing of 1.6 mm
and a diameter of 0.22 mm.  The geometry of the combs and the positions where
data was collected are shown on the right of Fig.\ \ref{fig:fouriers}.  This
particular orientation of the combs was found to produce the most isotropic and
homogeneous turbulence, with a nearly Gaussian velocity distribution and a
turbulence intensity low enough to rely on Taylor's frozen turbulence
assumption.

A laser velocimetry system (LDV) from TSI Inc. is used to make the velocity
measurements. Data is taken for two hours in the center of the film at each of
four vertical locations to obtain data sets of 33.6 million points each.
Roughly 14 kilometers of film flows past the measurement location during the
taking of each data set, with velocities being recorded every 0.4 mm on average
(with a typical data rate of 5kHz). The mean velocity of the film at location
(A) in Fig.\ \ref{fig:fouriers} is $\bar v = 214$ cm/s with a root mean square
deviation $\langle \Delta v^{2} \rangle ^{1/2} = 21$ cm/s (where $\Delta v = v
- \bar v$). The turbulence intensity is therefore $I_{t} = 9.8$\% at point A,
less than half of values reported in other work on 3D
turbulence\cite{Tabeling96a,Anselmet84a}. At locations progressively farther
below the combs we see a corresponding change in the magnitude of the
turbulence intensity, decreasing to 8.0\% at B, 6.8\% at C, and 6.0\% at D.

From the data sets, the $P(\delta v(r))$ were calculated at various values of
$r$, and from these, the functions $G_p(r)$ were calculated by numerical
integration akin to Eq.~\ref{eq:Sp}. It should be noted that there is an
inherent limitation on the maximum order of the structure function that can be
calculated in any physical experiment. As $p$ is increased the majority of the
weight of the integrand comes increasingly from the tails of the PDFs, where
the number of data points becomes quite small. We found that the majority of
the weight of the numerical integration of $G_p(r)$ was not contributed by the
wings of the PDFs for $p \leq 14$.

\begin{figure}
\includegraphics[width=3.2in]{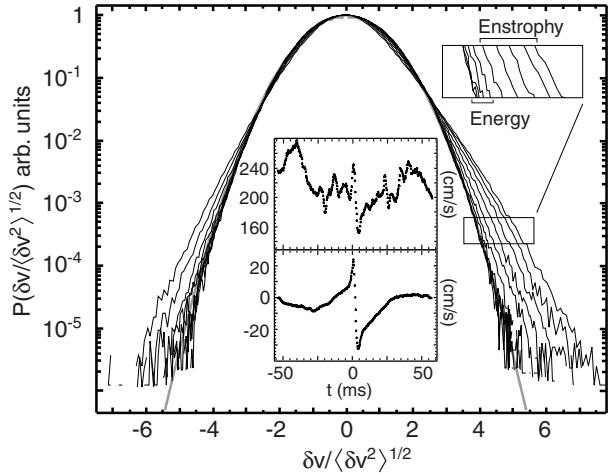}
\vspace{0.1in} \caption{Normalized velocity difference probability
distributions for a range of separations through the energy and enstrophy
cascades from data taken at location A. The thick gray line is a Gaussian
reference curve. The upper inset shows one particularly steep velocity
fluctuation event representative of the wings of the distributions. The lower
inset is an average of all such events with $| \delta v / \langle \delta v^2
\rangle^{1/2} | \geq 5.35$ at a separation of 0.5 cm (there were a total of
311). } \label{fig:dvpdf}
\end{figure}

Furthermore, as suggested by Anselmet {\sl et al.} \cite{Anselmet84a}, the
convergence of the structure functions with an increase in the size of the data
set provides a check on the quality and quantity of the data. If each small
subset of the data is statistically identical to all other subsets, we expect
to arrive at the same value for the $G_p(r)$ regardless of the size of the
subset used in the calculation or which subset in particular is chosen. We find
that for all separations $r$, the variation of $G_6(r)$  is below 5\% for data
sets of $5 \cdot 10^6$ points or greater.  For $G_{12}(r)$, $25 \cdot 10^6$
points are needed to meet the same requirement. Given the above considerations
we are confident of the measured structure functions up to order 12.

The energy spectrum in Fig.~\ref{fig:fouriers} shows simultaneous evidence of
the two scaling regimes, with $E(k) \propto k^{-5/3}$ in the inverse energy
cascade (associated with driven turbulence) and $E(k)\propto k^{-3}$ in the
forward enstrophy cascade (associated with decaying turbulence). The location
of the bend between the two scaling ranges gives an estimate of the effective
injection length scale at each location. We estimate it to be $k/2\pi =
1/\lambda = 0.63 \text{ cm}^{-1}$ or $\lambda = 1.6$ cm at location A.
Subsequent figures will refer to the energy and enstrophy cascades which are
inferred from the spectrum at A. More details on the turbulent spectra and the
nature of the injection scale were previously published by Rutgers
\cite{Rutgers98a}.

\begin{figure}
\includegraphics[width=3.2in]{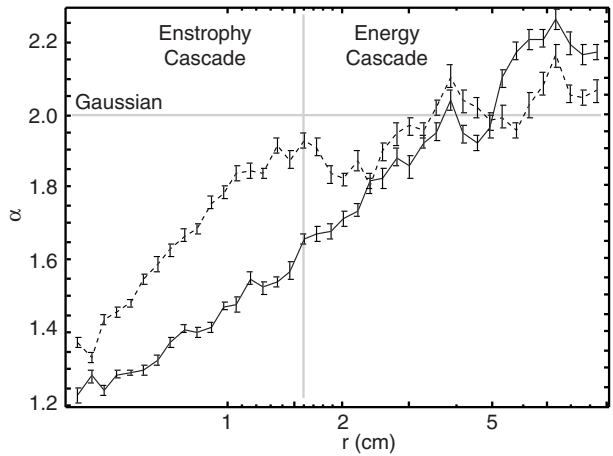}
\vspace{0.1in} \caption{The exponent $\alpha$ of the best fit stretched
exponential function $P(\delta v) \propto \exp{ \left\{ -|\delta v /
\sigma|^\alpha \right\}}$ (for data from location A). The solid line is the fit
for $\delta v / \langle \delta v^2 \rangle^{1/2} > 2$ (i.e. right wing); the
dashed line corresponds to $\delta v / \langle \delta v^2 \rangle^{1/2} < -2$
(left wing).} \label{fig:alphas}
\end{figure}

With the energy and enstrophy cascade inertial ranges defined, we inspect the
data for signs of intermittency. Recent experiments on the separate energy and
enstrophy cascades of a quasi 2D experimental system by Paret and
Tabeling\cite{Paret98a,Paret99a} found Gaussian velocity difference PDFs
independent of the scale and concluded that there was a complete absence of
intermittency. From the analysis of our forced turbulent data (measured at
point A), we conclude quite the opposite. Figure\ \ref{fig:dvpdf} shows
$P(\delta v(r))$ for $r$ covering both the forward enstrophy and inverse energy
cascades. To accentuate relative shape changes the width of each distribution
was rescaled by its second moment. The distributions with $r$ in the energy
cascade are close to Gaussian and differ substantially from distributions with
$r$ in the enstrophy cascade, which are decidedly more exponential in nature.
This variation of the shape can be quantified by fitting the distributions to a
stretched exponential function \cite{Tabeling96a,Kailasnath92a}, $P(\delta v)
\propto \exp{\{-|\frac{\delta v}{\sigma}|^{\alpha}\}}$.  The dependence of the
stretching exponent $\alpha$ on $r$ suggests a flow-field which is not
self-similar and is an indication of intermittent turbulent flow. The positive
and negative tails of the PDFs beyond $2 \sigma$ are then fit
independently\cite{Tabeling96a,Kailasnath92a}. The trend toward a slower than
Gaussian falloff at small length scales is shown in Fig.\ \ref{fig:alphas}. The
minimum $\alpha$ for negative $\delta v$ is greater than for positive $\delta
v$ reflecting the inherent asymmetry. Note that the sign of the asymmetry is in
keeping with the exact theoretical result, $S_3(r)=\frac{3}{2}\varepsilon r$.

The events in the raw velocity data which populate the wings of the $P(\delta
v(r))$ can be singled out. A typical event is shown in the upper inset of Fig.\
\ref{fig:dvpdf}.   The lower inset shows an average of all such events for
$r=0.5$ cm and $|\delta v(r) / \langle \delta v(r)^2 \rangle ^{1/2}| \geq 5.35$
(an arbitrary cutoff). Since the third moment of $P(\delta v(r))$ is positive,
the average has a corresponding characteristic sharp negative slope.  Note that
the insets in Fig.~\ref{fig:dvpdf} have been plotted as raw time traces from
the measurement probe, as is customarily done in previous work.  The
application of the frozen turbulence assumption will lead to a shock with
positive slope which gives a positive $S_3(r)$ for our definition of $\delta
v(r)$. The shock-like nature of these events bears a strong qualitative
resemblance to intermittent events measured in 3D turbulence, but typically
with a velocity increment of opposite sign\cite{Belin96a,Willaime98a}, since
$S_3(r)$ is negative for 3D turbulence.

\begin{figure}
\includegraphics[width=3.2in]{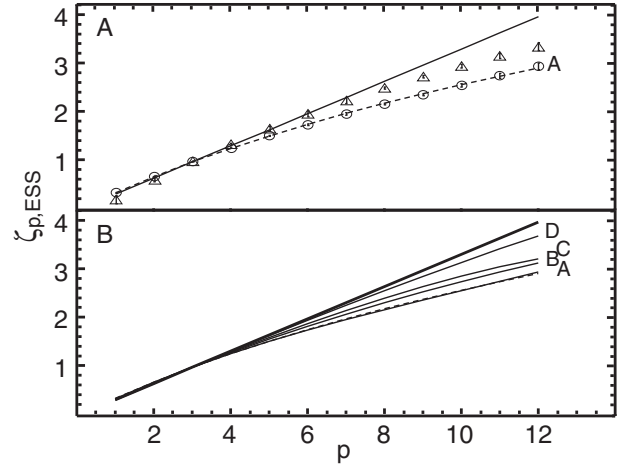}
\vspace{0.1in}\caption{{\bf A}. ESS scaling exponents in the inverse energy
$(\bigtriangleup)$ and forward enstrophy $(\bigcirc)$ cascades at location (A)
(see Fig.\ \ref{fig:fouriers}). The solid line is the K41 prediction, and the
dashed line the log-Poisson. Error bars show the 95\% confidence interval. {\bf
B}. Measurements of $\zeta_{p,{\rm ESS}}$ in the enstrophy range at points
A,B,C, and D (see Fig.\ \ref{fig:fouriers}) } \label{fig:scaling}
\end{figure}

These intermittent events have a strong effect on the structure functions
$S_p(r)$, which theoretically scale with $r$.  Practically, it is extremely
difficult to discern this scaling directly since one typically requires several
decades of clearly discernible scaling in $E(k)$.  Most experiments, ours
included, do not have that luxury. The standard remedy follows a technique
introduced by Benzi {\sl et al.} \cite{Benzi93a} where $S_p(r)$ is plotted
against $G_3(r)$, leading to much cleaner power laws throughout the inertial
range. This procedure is referred to as extended self-similarity and is used in
most studies of intermittency in 2D and 3D turbulence. We will refer to the
scaling exponents derived from this technique as $\zeta_{p,{\rm ESS}}$.
Figure~\ref{fig:scaling}A shows the $\zeta_{p,{\rm ESS}}$ as a function of the
order $p$, for data collected at point A. The solid line is the prediction in
the absence of intermittency, and the dashed curve the log-Poisson
prediction\cite{She94a} (which is a good fit to the canonical 3D intermittency
measurements of Anselmet {\sl et al.}\cite{Anselmet84a}). Our scaling exponents
from the enstrophy cascade of forced 2D turbulence ($\bigcirc$) are nearly
indistinguishable from those for 3D measurements. Exponents from the energy
cascade deviate as well from the self-similar prediction, but to a lesser
degree ($\bigtriangleup$).

At first sight our measurements appear to contradict those of Paret and
Tabeling mentioned earlier\cite{Paret98a,Paret99a}. There is, however, an
important difference between our measurements and theirs. Paret and Tabeling's
electroconvection cell is limited by its range of accessible length scales and
is capable of producing only the inverse energy cascade or the forward
enstrophy cascade independently. Computer simulations are often similar in
scope when they employ hyperviscous terms to suppress the enstrophy range when
the energy range is being studied \cite{Smith94a}. The results shown in Fig.\
\ref{fig:scaling} are for a system with {\sl both} cascades present
simultaneously.

To verify the importance of two cascades versus one, we performed the analysis
leading to Fig.\ \ref{fig:scaling}A again for points progressively further
downstream from the combs where the turbulence has decayed (points B, C, and D
in Fig.\ \ref{fig:fouriers}). The results are summarized in Fig.\
\ref{fig:scaling}B. At the largest distance below the comb (D) the spectrum
indicates only an enstrophy cascade\cite{Rutgers98a,Batchelor69a,Chasnov97a}.
Accordingly the indications of intermittency disappear with $\zeta_{p,{\rm
ESS}}$ tending toward $p/3$. Our results at this location thus agree with the
electroconvection experiments\cite{Paret99a} on the isolated enstrophy cascade.
We were not able to produce the case of an isolated energy cascade in the
flowing soap films and thus, cannot make a comparison to the corresponding
electroconvection experiments\cite{Paret98a} which also found no intermittency.
Soap film and electroconvection experiments therefore agree for similar
situations, but both disagree with parts of simulations by Babiano {\sl et
al.}, which claim intermittency for inverse energy cascades but never for
forward enstrophy cascades, regardless of whether the cascades are isolated or
simultaneous.

There is always a question of the viability of flowing soap films as effective
2D incompressible fluids. It is possible that our results are influenced by
quantities specific to flowing soap films, such as air drag, film thickness
variations, or compressibility effects.  Though these effects are important and
we will make them the subject of further study, the clear identification of the
energy and enstrophy cascades in flowing films \cite{Rutgers98a,Martin98a}
justifies this experimental medium. To this can be added our current agreement
with electroconvection experiments \cite{Paret98a,Paret99a} that there is no
significant intermittency in 2D turbulent flows with an isolated enstrophy
cascade.

In conclusion, find evidence of intermittency in forced 2D turbulence when both
the inverse energy and forward enstrophy cascades are present simultaneously.
Signs of intermittency are strongest in the enstrophy cascade, and present to a
lesser degree in the energy cascade.  Evidence of intermittency is threefold.
First we find a considerable dependence of the shape of the velocity difference
PDFs on the scale $r$.  As $r$ is increased through the enstrophy and energy
cascade ranges the tails of $P(\delta v(r))$ can be fit by a stretched
exponential with exponent $\alpha$ varying from 1.23 toward Gaussian
statistics. Secondly we can identify shock like changes in the velocity which
contribute to the tails of $P(\delta v(r))$. Thirdly, there is a deviation of
the ESS structure function scaling exponents from the K41 prediction in both
the energy and enstrophy cascades. The experimental values lie well-below the
K41 prediction of $\zeta_p=p/3$ in the energy cascade, and in the enstrophy
cascade lie directly on top of the log-Poisson prediction\cite{She94a}. The
same system shows none of these signs of intermittency when the turbulence is
left to decay, in which case there is only an isolated enstrophy cascade. This
observation agrees with other experiments using an electronconvection
technique\cite{Paret99a}.  The simultaneous presence of both inverse energy and
forward enstrophy cascades appears necessary to observe intermittency in 2D
turbulence.

\acknowledgements{We would like to thank C. Jayaprakash and F. Hayot for their
stimulating discussions. The present work was supported by the Petroleum
Research Foundation and The Ohio State University.}

\end{multicols}

\end{document}